\documentclass[twocolumn,showpacs,preprintnumbers,amsmath,amssymb]{revtex4}

\usepackage{graphicx}
\usepackage{dcolumn}
\usepackage{bm}

\begin{document}


\title{Spatially dependent Rabi oscillations: \\an approach to sub-diffraction-limited CARS microscopy}

\author{Willem P. Beeker}
\author{Chris J. Lee} 
\author{Klaus-Jochen Boller}  
 \affiliation{Laser Physics \& Nonlinear Optics Group, MESA+ Research Institute for Nanotechnology, 
University of Twente, P. O. Box 217, Enschede 7500AE, The Netherlands
}
\email{k.j.boller@utwente.nl}
\homepage{http://lpno.tnw.utwente.nl/}

\author{Petra Gro\ss}
\author{Carsten Cleff}
\author{Carsten Fallnich}
\affiliation{Institut f\"{u}r Angewandte Physik, Westf\"{a}lische Wilhelms-Universit\"{a}t, 48149 M\"{u}nster, Germany}

\author{Herman L. Offerhaus}
\author{Jennifer L. Herek}
\affiliation{Optical Sciences Group, MESA+ Research Institute for Nanotechnology, 
University of Twente, P. O. Box 217, Enschede 7500AE, The Netherlands}
\date{\today}

\begin{abstract}
We present a theoretical investigation of coherent anti-Stokes Raman scattering (CARS) that is modulated by periodically depleting the ground state population through Rabi oscillations driven by an additional control laser. We find that such a process generates optical sidebands in the CARS spectrum and that the frequency of the sidebands depends on the intensity of the control laser light field. We show that analyzing the sideband frequency upon scanning the beams across the sample allows one to spatially resolve emitter positions where a spatial resolution of 65~nm, which is well below the diffraction-limit, can be obtained.
\end{abstract}

\pacs{31.15.ap, 87.64.-t, 78.20.Bh, 78.47.Fg}
                             
\maketitle

\section{introduction}

Optical microscopy is one of the key techniques used to analyze biological processes at systemic, cellular, and sub-cellular levels. However, current optical microscopy techniques have insufficient resolution to observe interactions at the resolution desired by cell biologists--that is at the molecular or single functional group level.
 
Current near-field techniques are sub-diffraction-limited, such as scanning near-field optical microscopy \cite{heinzelmann} but they suffer from the disadvantage that they are limited to surfaces. Two notable far-field techniques have achieved sub-diffraction-limited resolution: stimulated emission depletion (STED) microscopy \cite{hell}, and stochastic optical reconstruction microscopy (STORM) \cite{rust,huang}. Both techniques require that the sample is labeled with fluorescent dyes. In addition, STED requires that the intensity at the sample is rather high to resonantly saturate the electronic transitions of the label, hastening photo-bleaching and potential cyto-toxicity \cite{ikagawa,szeto}. STORM, on the other hand, uses very low light intensities, but requires very special labels that can be intentionally switched between dark and light states. Furthermore, STORM images take a substantial amount of time to acquire \cite{rust}, obstructing live cell imaging.

Ideally, biologists would prefer to employ label-free imaging techniques, such as CARS microscopy, which replaces labeling by detecting the chemically specific vibrational modes of the molecules. Furthermore, the non-resonant nature of the excitation process limits photo-damage \cite{fu}. Techniques to achieve sub-diffraction-limited CARS imaging are the subject of active investigation. Linear techniques to suppress CARS emission have been the subject of both experimental and theoretical investigation \cite{nikolaenko}. However, linear techniques cannot be used to obtain sub-diffraction-limited resolution. In previous work, we investigated a saturation process, analogous to STED, that prevents the build-up of the vibrational coherence required for CARS emission \cite{beeker}. However, the saturation was presumed to be fully incoherent, via a combination of long lifetimes and short dephasing times, leading us to explore alternative approaches that also include coherent effects.

\begin{figure}
\includegraphics[width=60mm]{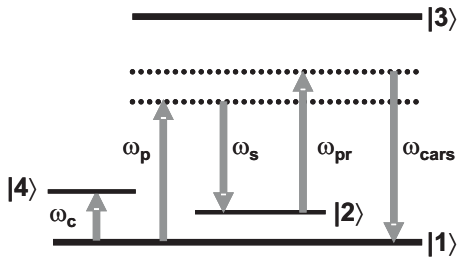}
\caption{\label{fig:levelscheme} Energy level diagram for the CARS process with an additional level $\left|4\right\rangle$. Level $\left|1\right\rangle$ is the ground level and initially fully occupied. Level $\left|2\right\rangle$ is a vibrational level of the medium, $\left|3\right\rangle$ is the excited level and level $\left|4\right\rangle$ the control state, which has a low decoherence rate. The arrows indicate possible transitions induced by the incident laser fields, which are far detuned from the transition frequency ($\omega_{p}$, $\omega_{S}$ and $\omega_{pr}$), or on resonance ($\omega_{c}$). Through coherent anti-Stokes Raman scattering (CARS), the medium emits light at the frequency, $\omega_{CARS}$, while the additional control field, $\omega_{c}$, drives Rabi oscillations at the $\left|1\right\rangle$~-~$\left|4\right\rangle$ transition.}
\end{figure}

Here, we use a density matrix approach \cite{milonni} to model the CARS emission process. We consider CARS emission in a four-level system (see Fig.~\ref{fig:levelscheme}) in combination with an additional light field that couples an additional level $\left|4\right\rangle$, called control level, to the ground-state level $\left|1\right\rangle$. A mechanism that leads to sub-diffraction-limited CARS microscopy has been identified for this system: for the case that the lifetime of $\left|4\right\rangle$ and the dephasing of the the $\left|1\right\rangle$-$\left|4\right\rangle$ transition are long compared to the inverse transition rate imposed by the additional laser, Rabi oscillations between $\left|1\right\rangle$ and $\left|4\right\rangle$ induce a Rabi-splitting of the CARS emission. The Rabi-splitting is intensity-dependent, which can be used to resolve features within a diffraction-limited volume. 

\section{Theoretical framework}

A detailed description of the level scheme, density matrix equations, and light fields has been given elsewhere~\cite{beeker}. In summary, the four-level system (see Fig.~\ref{fig:levelscheme}) has a ground state ($\left|1\right\rangle$), a vibrational state ($\left|2\right\rangle$), an excited electronic state ($\left|3\right\rangle$), and the control state ($\left|4\right\rangle$). Transitions between the ground and excited states, the vibrational and excited states as well as the ground and control states are dipole allowed, while all other transitions are dipole forbidden. The medium is irradiated with four pulsed light fields, two of which (pump, $\omega_{p}$; Stokes, $\omega_{S}$) are two-photon resonant with the $\left|1\right\rangle$~-~$\left|3\right\rangle$ transition, driving population into $\left|2\right\rangle$ and inducing a coherence between $\left|1\right\rangle$ and $\left|2\right\rangle$ (we refer to this as the vibrational coherence). The vibrational coherence induces two sidebands on the probe light field ($\omega_{pr}$), one of which is known as the CARS emission frequency ($\omega_{p}~-~\omega_{S}~+~\omega_{pr}~=~\omega_{cars}$).

The additional laser (called the control laser $\omega_{c}$), resonant with the $\left|1\right\rangle$~-~$\left|4\right\rangle$ transition, influences the CARS emission. To understand the control laser's effect, we calculate the temporal evolution of the laser pulse envelopes and the envelopes of the density matrix elements, along with their radiating fields, as described in detail in reference \cite{beeker}. We study CARS emission for the case where $\left|4\right\rangle$ has a decoherence rate that is slow compared to the duration of the laser pulses, resulting in Rabi oscillations between $\left|1\right\rangle$ and $\left|4\right\rangle$. Although $\left|4\right\rangle$ is considered to be a vibrational state in this calculation such as in \cite{witte} or \cite{ventalon}, any dipole allowed state with the requisite decoherence rate can be used.

For generality, we choose energy levels, detunings and pulse durations that are typical for CARS emission processes from molecules. The $\left|1\right\rangle$~-~$\left|3\right\rangle$ transition angular frequency is set to 4,700~THz (approx.~400~nm or 25,000~cm$^{-1}$), and the $\left|1\right\rangle$~-~$\left|2\right\rangle$ frequency to 314~THz (approx.~6~$\mu$m or 1,667~cm$^{-1}$). Likewise, we choose a $\left|1\right\rangle$~-~$\left|4\right\rangle$ transition frequency of 565~THz (approx.~3.33~$\mu$m or 3,000~cm$^{-1}$).

The combination of the pump laser wavelength as 800~nm, and Stokes as 923~nm is taken to provide two-photon resonance with the $\left|1\right\rangle$~-~$\left|2\right\rangle$ transition. The probe laser wavelength is taken as 600~nm. The total lifetime of state $\left|3\right\rangle$ and the lifetime of state $\left|4\right\rangle$ are taken to be on the order of nanoseconds \cite{vivie}, while the decoherence rates between states involved in the CARS emission process are of the order of picoseconds (see e.g., \cite{vivie,asbury,wurzer}). All the laser pulse durations $\tau$ are set to 7~ps~(1/$e^2$) except for the control laser field, which is continuous during the interval of CARS emission. The numerical calculations extend over 30~ps in steps of 0.1 femtoseconds. The amount of output data was set to 70,000 points for fast computation and causes only a small amount of discretization noise. 

\section{results}
\begin{figure}
\includegraphics[width=70mm]{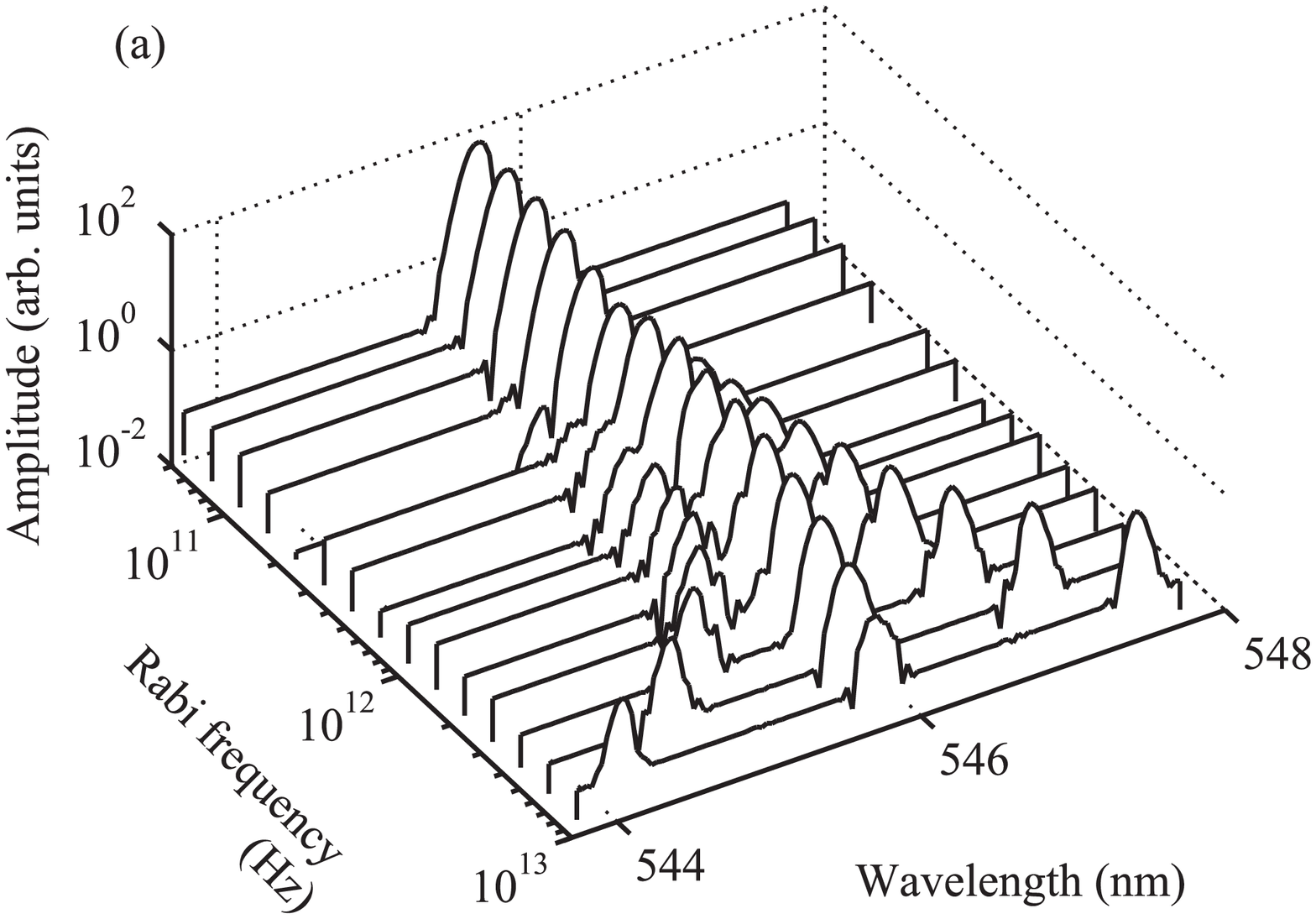}
\includegraphics[width=70mm]{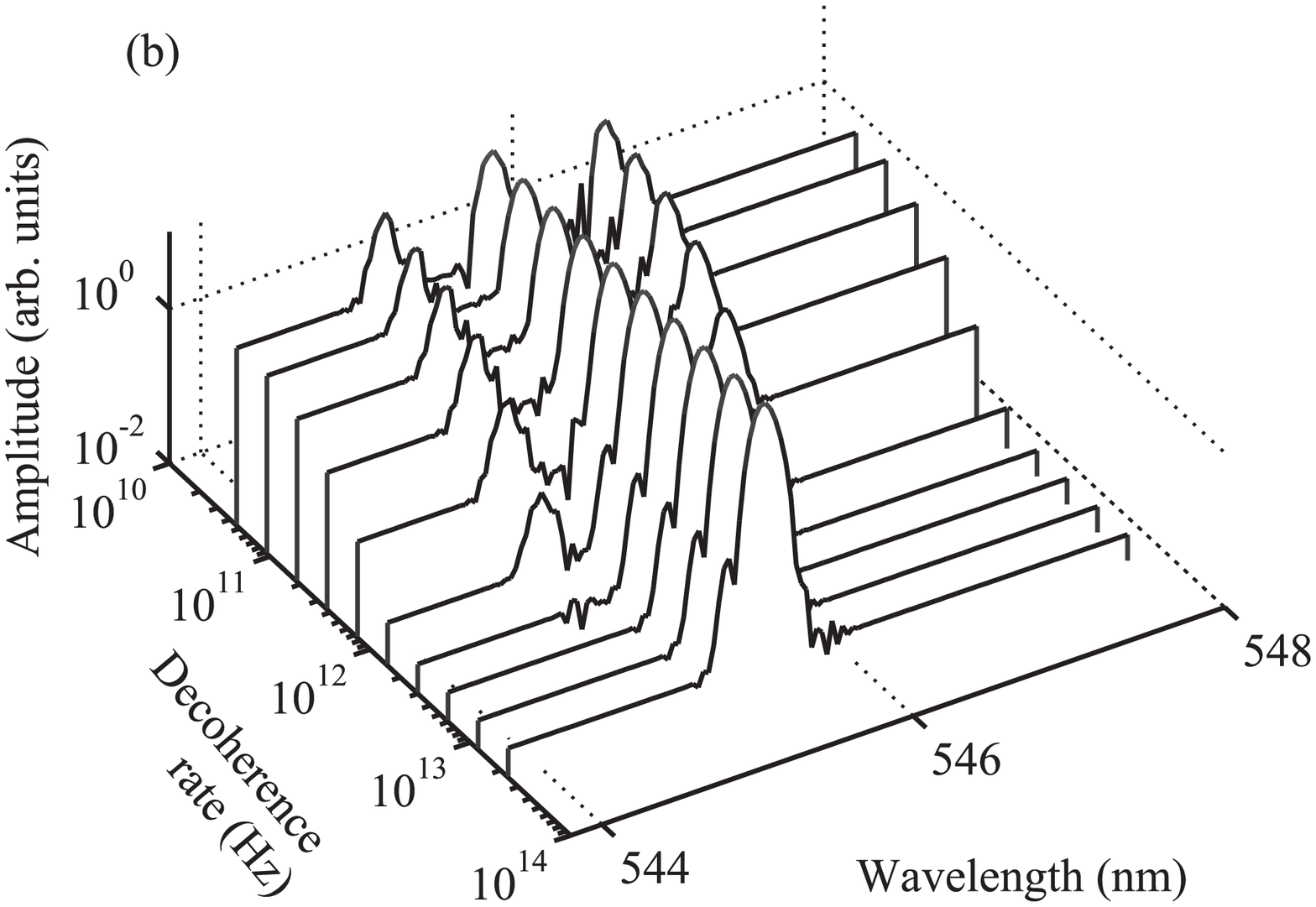}
\caption{(a) Emission spectra containing the CARS emission peak for increasing values, on a logarithmic scale, of the induced Rabi frequency oscillations between states $\left|1\right\rangle$ and $\left|4\right\rangle$. The amount of splitting of the sidebands is equal to $\Omega_{R}$. (b) Emission spectra containing the CARS emission peak for increasing values, on a logarithmic scale, of the decoherence rate of $\Gamma_{14}$ with $\Omega_{R}$~=~4~THz for each spectrum. Note that the amplitude of the sidebands decreases and disappear around where $\Omega_{R}$ equals the decoherence rate. The increase in Rayleigh scattering of the incident light fields for the lower set of decoherence rates, leads to an increase in background signal at the CARS emission wavelength.}
\label{fig:decoherence}
\end{figure}

We performed a series of calculations of the emission spectrum around the CARS frequency. In these calculations $\omega_c$ was taken to be resonant with the $\left|1\right\rangle$~-~$\left|4\right\rangle$ transition (the detuning $\Delta_{14}=0$), while the decoherence rate of this transition was set to $\Gamma_{14} = 0.1$~THz. The amplitude of the control field and, hence, the Rabi frequency ($\Omega_R$) were varied in steps. The results are shown in Fig.~\ref{fig:decoherence} where the spectrum around the CARS frequency, as emitted by the medium, for a number of values of $\Omega_R$ is given. A single emission peak is seen at low Rabi frequencies and indicates standard CARS emission, as expected. It can be seen that, for increasing $\Omega_R$, the spectrum shows two symmetrically placed sidebands. From our calculations, we find that these sidebands show a spacing from the CARS carrier frequency that coincides with $\Omega_R$, given by \cite{milonni}: 

\begin{eqnarray}
\Omega_{R}=\sqrt{\frac{E_c^2\mu^{2}_{14}}{\hbar^{2}}+\Delta^{2}_{14}}
\label{eq:rabifrequency}
\end{eqnarray}

Here $E_c = \sqrt{(2I_c/\epsilon_0)}$ is the electric field of the control laser and $I_c$ the control intensity, and $\mu_{14}$ is the dipole moment at the control transition. From this, we conclude that the Rabi oscillation periodically depletes the ground state, which, in turn, modulates the vibrational coherence. As a result, the CARS emission becomes amplitude modulated which spectrally shows up as two Rabi sidebands. The splitting of several THz is large enough to be distinguishable in a real CARS experiment using standard spectrum analyzers. 

It is critical, for obtaining a noticeable Rabi splitting in the CARS spectrum, that the frequency of the Rabi oscillations exceeds the decoherence rate $\Gamma_{14}$, as can be seen in Fig.~\ref{fig:decoherence}(b), where the spectrum of the CARS emission, modulated with $\Omega_R=4$~THz, is shown for increasing values of the decoherence rate. Up to decoherence rates which correspond to sub-picosecond decoherence times, the 4~THz modulation is still detectable in Fig.~\ref{fig:decoherence}(b). Obtaining such high Rabi frequencies is important, because the decoherence times in liquids are very short, typically on the order of 5~ps~\cite{witte}.

The induced sidebands can be used to obtain positional information by noting from Eq.~\ref{eq:rabifrequency} that the Rabi splitting of the sidebands depends on the intensity of the control laser. Furthermore, the geometrical alignment of the dipole moment $\mu_{14}$ may be obtained by rotation of the polarization angle of the control laser beam, providing additional information on the structure of the sample. Considering a single emitter illuminated by a Gaussian beam, we note that the radial distance of the emitter from the center of the control beam can be obtained by measuring the exact frequency of the sidebands, which can be called a Rabi-labeling of the emitter position. Such a space-dependent Rabi frequency is a well-known effect \cite{schouwink}, however, it is usually considered to cause undesired broadening and is eliminated, e.g., in pump-probe experiments, by using a much larger pump beam as compared to the probe beam \cite{sautenkov}.

The absolute position of an emitter location can be calculated by trilateration. We illustrate this with a two dimensional example, however, a similar approach also applies to three dimensions. Consider an emitter located at the Cartesian coordinates $x_{1}$ and $y_{1}$. The observed CARS emission spectrum, with its sideband frequency depending on the local intensity of the control laser, will not determine these coordinates but instead provides the distance from the control beam center, $r_1=\sqrt{x_1^2+y_1^2}$, which is a ring of possible locations centered around the control beam. By scanning the control laser a known distance $dx$ along the $x$ axis, the radial location of the emitter is changed to $x_{1}$~-~$dx$, $y_{1}$. As a result, the emitter modulation frequency will change, revealing a new distance, $r_2=\sqrt{(x_1-dx)^2+y_1^2}$. The expressions for $r_1$ and $r_2$ can then be solved for $x_1$, and a subsequent scan along $y$ by a known $dy$ determines $y_1$ as well. Turning our attention to the spatial resolution of such a positioning scheme, we show that the scheme can provide CARS with sub-diffraction-limited resolution.

We note that the control laser, resonant with the $\left|1\right\rangle$~-~$\left|4\right\rangle$ transition and focused to a diffraction-limited Gaussian-shaped intensity profile, centered on the other laser beams, illuminating a distribution of emitters, will generate a distribution of Rabi sidebands on the CARS signal. If the control laser has a peak envelope field strength of $E_{0}$ and a 1/$e$-spot-size of $w_{0}$, then the range of observed Rabi frequencies within the diffraction-limited spot is $\Omega_{max}=\left|\mu_{14}E_{0}/\hbar\right|$ in the center to $\Omega_{min}=1/e\left|\mu_{14}E_{0}/\hbar\right|$ at radial distance $w_{0}$. The radial resolution is then determined by how accurately the Rabi splitting frequency can be measured in the CARS spectra. Given a Rabi frequency, $\Omega$, corresponding to a radial distance from the center of the focus of $r$, and a frequency measurement accuracy of $d\Omega$, corresponding to a radial resolution of $dr$, it can be shown that
 
\begin{eqnarray}
dr = \frac{w_0^2}{2r+dr}\ln\left(\frac{\Omega}{\Omega-d\Omega}\right)
\label{eq:two}
\end{eqnarray}

\begin{figure}
\includegraphics[width=70mm]{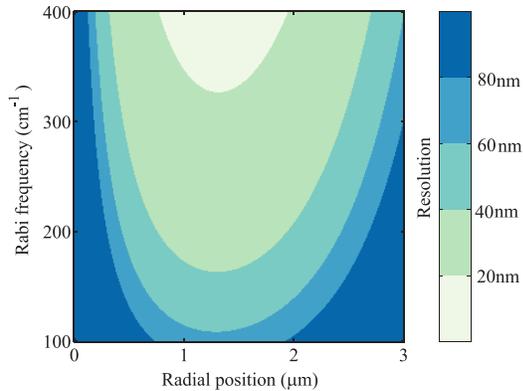}
\caption{Spatial resolution as a function of the Rabi frequency and radial location of the emitter.}
\label{fig:resolutionimprovement}
\end{figure}

It can be seen from the transcendental equation~\ref{eq:two} that $dr$ becomes smaller than the diffraction-limit (smaller than $w_0$), and that higher Rabi frequencies lead to higher resolution, while emitters located away from the center of the control beam are resolved better than emitters close to the center. The resolution is expected to be maximal at the steepest intensity slope of the control beam, therefore, by using a Gaussian shaped control beam, the highest resolution is expected to be found off center. This is illustrated in Fig.~\ref{fig:resolutionimprovement} where we have plotted the Rabi-labeling resolution as a function of both Rabi frequency and radial location, for $d\Omega=3$~cm$^{-1}$ (which is a typical value for the linewidth of the CARS peak in a spectrum). This figure shows that considerable improvement over diffraction-limited CARS microscopy is possible. 

To make the improvement explicit, we consider a specific example for a microscope with a numerical aperture of 1.2 and laser wavelengths that were used in our calculations. In this case, the diffraction-limited resolution is 171~nm. If the control laser ($\lambda$=3.3~$\mu$m, focused to a diffraction-limited spot, induces 100~cm$^{-1}$ Rabi oscillations, then at a radial distance of 1.2~$\mu$m, the resolution is 65~nm. We estimate from the data presented in ~\cite{witte} that the laser intensity required to generate a Rabi frequency of 100~cm$^{-1}$ is about 500~MW/cm$^{2}$, which is comparable to the intensities used in standard CARS experiments, and sufficiently low to avoid multiphoton excitation \cite{ventalon}. However, it should be noted that such a value will depend on the molecular system under study.

\section{conclusions}

In conclusion, we have demonstrated a route to obtaining sub-diffraction-limited CARS signals. By resonantly driving the excitation of a control state with relatively long decoherence times, intensity and thus spatially dependent sidebands in the CARS emission spectrum can be generated. By accurately measuring this Rabi splitting in the CARS spectrum, objects can be resolved to a resolution below the diffraction-limit. Using typical numbers, we show that a resolution in the order of 65~nm may be achievable. We note that our calculations also show the generation of Rabi sidebands in other molecule-specific emission lines, such as Coherent Stokes Raman Scattering. The approach towards sub-diffraction-limited resolution presented here can thus be used with this microscopy technique as well, making it more broadly applicable.

\bibliography{CARSlibRabi}
\bibliographystyle{apsrev}
\end{document}